\documentclass[12pt]{iopart}
\usepackage{graphicx}
\usepackage{epsfig}
\begin{document}

\title[Oscillatory processes in solar flares]{Oscillatory processes in solar flares}

\author{V~M~Nakariakov$^1$,  A~R~Inglis$^2$, I~V~Zimovets$^3$,  C~Foullon$^1$,  E~Verwichte$^1$, R~Sych$^4$ and I~N~Myagkova$^5$}

\address{$^1$ Physics Department, University of Warwick, Coventry, CV4 7AL, UK}
\address{$^2$ School of Mathematics and Statistics, University of St Andrews, St Andrews, KY16 9SS, UK}
\address{$^3$ Space Research Institute, Russian Academy of Sciences, Profsoyuznaya St. 84/32, Moscow, 117997, Russia}
\address{$^4$ Institute of Solar-Terrestrial Physics, Lermontov St., 126a, Irkutsk, 664033, Russia}
\address{$^5$ Skobeltsyn Institute of Nuclear Physics, Lomonosov Moscow State University, Moscow 119992, Russia}
\ead{V.Nakariakov@warwick.ac.uk}

\begin{abstract}
Electromagnetic (radio, visible-light, UV, EUV, X-ray and gamma-ray) emission generated by solar and stellar flares often contains pronounced quasi-periodic pulsations (QPP). Physical mechanisms responsible for the generation of long-period QPP (with the periods longer than one second) are likely to be associated with MHD processes. 
The observed modulation depths, periods and anharmonicity of QPP suggest that they can be linked with some kind of MHD auto-oscillations, e.g. an oscillatory regime of magnetic reconnection. Such regimes, of both spontaneous and induced nature, have been observed in resistive-MHD numerical simulations. The oscillations
are essentially nonlinear and non-stationary. We demonstrate that a promising novel method for their analysis is the Empirical Mode Decomposition technique.
\end{abstract}
 
\vspace{2pc} 
\submitto{\PPCF}  

\section{Introduction}

Solar flares are sudden releases of magnetic energy
which occur in the solar atmosphere, reaching up to 10$^{32}$ ergs and
lasting anything from a few to a few tens of minutes (see, e.g., \cite{2008LRSP....5....1B} 
for a recent comprehensive review). 
Flare emission is detected in all EM bands, from radio to gamma-rays, and also are
detected in the variations of the flux density of solar energetic particles approaching the Earth
with energies ranging from a few tens of 
keV to several GeV. Similar
events are observed on solar type stars, as well as on stars of other classes, e.g. of the UV Cet type.
Solar flares, being among the most powerful and most energetic physical 
phenomena in the solar system, attract attention because of their decisive role in solar-terrestrial
relations and space weather, their connection with the Earth's climate, as well as in the context of
basic plasma physics research.

Often, the EM radiation generated in solar and stellar flares shows a pronounced oscillatory
pattern, with characteristic periods ranging from a fraction of a second to several minutes. The oscillations
can be seen in all observational bands, often in phase, and the modulation depth sometimes exceeds
100\% (see, e.g. \cite{2009SSRv..149..119N}). Traditionally, these oscillations are referred to as
quasi-periodic pulsations (QPP), to emphasise that they often contain apparent amplitude and
period modulation, and that their shape is often anharmonic. 

Interest in solar flare QPP is connected first of all with the possible role these oscillations
play in flare energy releases. Indeed, there must be some physical reason for the flaring emission being
arranged in a sequence of periodic bursts in a significant fraction of observed flares. The 
occurrence of QPP put additional constraints on the interpretation and understanding 
of the basic processes operating in flares, such as particle acceleration, 
magnetic energy liberation and plasma hydrodynamics and thermodynamics.

Also, observed parameters of the oscillation (e.g. the periods, modulation, typical signatures, spatial 
information) should be
connected with the physical parameters of the flaring plasma and hence can be used for diagnostic
purposes through the method of coronal seismology (see \cite{2009PPCF...51l4019V} for more detail).
The QPP-based seismology has a number of attractive advantages. The high level of the emitted power 
allows observers to increase the time resolution of their instruments, easily resolving e.g. 
the Alfv\'en wave transverse transit time and perhaps even the ion gyroperiod. The flare excites various kinds of 
magnetohydrodynamic (MHD) waves and oscillations in the compressible and elastic surrounding plasmas, 
which can be detected in the modulation of the flaring emission. Moreover, QPP are also observed in stellar flares, and this opens up interesting perspectives for stellar coronal seismology and various comparative studies.

\begin{figure}
  \begin{center}
  \includegraphics[scale=0.55]{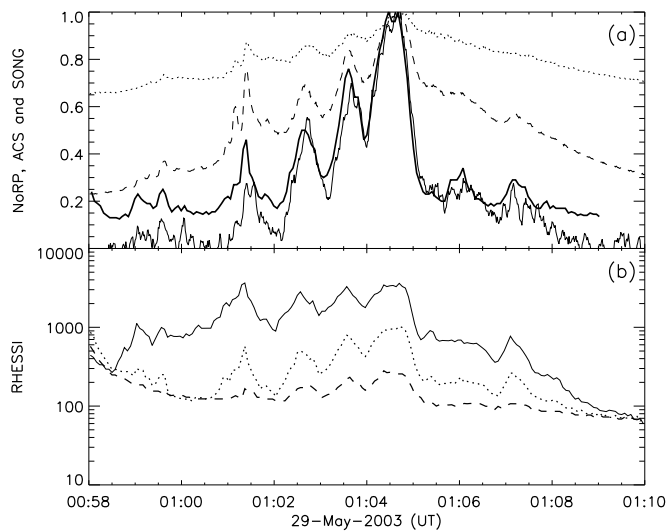}
  \end{center}
  \caption{A typical QPP in a solar flare. The event on 29 May 2003 observed with the Nobeyama
 Radiopolarimeter (NoRP) in microwaves, and in hard X-rays by the RHESSI spacecraft, the
 Anti-Coincidence Shield (ACS) of the SPI spectrometer on INTEGRAL, and 
the SOlar Neutrons and Gamma-rays experiment (SONG) on CORONAS-F.
Panel (a): NoRP flux at 17~GHz (dashed curve) and 35~GHz (dotted curve), the ACS count rate ($>80$ keV,
thin solid curve) and the SONG 80-200~keV count rate (thick solid curve). 
Panel (b): the RHESSI corrected count rate in the channels 25-50~keV (solid), 50-100
keV (dotted), and 100-300~keV (dashed).
}
\label{fig:l}
\end{figure}

Mathematically, in a number of cases, flaring QPP can be considered as an auto-oscillation:
a undamped oscillation in a non-linear dynamical system, whose amplitude and 
frequency are largely independent of the initial conditions, 
and are determined by the properties of the system itself. Dynamical systems capable of performing 
auto-oscillations include clocks, generators of electric vibrations, wind and string musical instruments, etc.
Many laboratory and natural plasma systems are observed to exhibit self-sustaining
oscillatory phenomena, which can be classified as auto-oscillations. 
Plasma physics examples of auto-oscillations are the sawtooth oscillations
\cite{1997Ap&SS.256..177H}, periodic shedding of hydrodynamic or
Alfv\'enic vortices in the interaction of magnetohydrodynamic flows with stationary obstacles 
\cite{2009A&A...502..661N},  oscillations of the magnetic reconnection rate and the current channel in laboratory 
experiments (e.g. \cite{2005PhPl...12e2107E}), and nonlinear thermal
over-stability of the plasma \cite{2010PhPl...17c2107C}. Investigation of plasma auto-oscillations is
a rapidly developing research topic, and the study of QPP in solar flares provide us with unique and
abundant information of fundamental importance.   

Here we present a brief review of the current trends in the observational and theoretical study of QPP in
solar flares, concentrating on their auto-oscillatory nature. Comprehensive discussion of this topic, including 
alternative theories, such as the modification of the non-thermal particle dynamics by a sausage mode
and oscillatory regimes of wave-particle interaction, can be found in \cite{2009SSRv..149..119N,1987SoPh..111..113A,2002SSRv..101....1A,2008PhyU...51.1123Z}.

\section{Observational example}
\label{sec:obs}

Often, QPP can be clearly seen in the time series (\lq\lq light curves") obtained in the observational channels, 
associated with non-thermal electrons, accelerated in the flare. Figure~\ref{fig:l} shows a typical example of light curves obtained in the microwave and hard X-ray bands
with four different instruments during a powerful solar flare (of the GOES-class X1.2). The independent detection
of QPP with different ground-based and space-borne instruments eliminates the possibility that
they are non-solar in origin. A detailed phenomenological study of this QPP event was recently presented in 
\cite{2009SoPh..258...69Z}. At least four one-minute QPP of growing amplitude are visible 
between 01:01 and 01:05 UT. A number of similar events are summarised in the on-line archive 
\textsf{http://www.warwick.ac.uk/go/cfsa/people/valery/research/qpp/} 
which provides a list of currently known solar QPP events. 
An interesting feature of the QPP is its apparent anharmonic shape (see Section~\ref{sec:emd}).
Recently, it was discovered that in some QPP several different oscillatory patterns 
occur simultaneously \cite{2009A&A...493..259I}.

In weak flares, QPP can be seen in, e.g., soft X-ray emission \cite{Foullon10}. 
However, in strong flares, QPP are observed at photon energies up to 2-6 MeV, which are associated with 
accelerated ions \cite{2010ApJ...708L..47N}, as well as at the lower energies associated with
non-thermal electrons. This indicates that the likely cause of the QPP is the time variability of the
charged particle acceleration process, e.g. a periodic regime of magnetic reconnection.

\section{Oscillatory regimes of magnetic reconnection}

The mechanism responsible for the fast release of magnetic energy in the solar corona, and its conversion
into heat, and kinetic energy of bulk flows and accelerated charged particles is believed to be magnetic reconnection.
Details of this mechanism are still under intensive investigation; although it is clear that the process is
essentially non-stationary: the steady supply of the magnetic energy by reconnection inflows results in bursty
energy release. There are two possibilities for this process to be periodic:
spontaneous, when the periodicity is determined by the plasma parameters in the reconnection site,
and periodically triggered, when the periodicity is prescribed externally, e.g. by an external resonator.

\subsection{Spontaneous reconnection} 
\label{sec:spon}

Spontaneous periodic or quasi-periodic release of energy by magnetic reconnection can be considered as a 
load/unload 
or relaxation process. We may illustrate it by a \lq\lq dripping" model \cite{2009SSRv..149..119N}:  
slowly and continuously leaking water is gathering on the ceiling, forming a bulge of growing mass, and when the 
gravitational force becomes sufficiently strong to counteract the surface tension force, the droplet reconnects 
from the rest of the water and falls down. Then the situation repeats, and the dripping rate can be quite 
periodic and stable. The 
period of dripping is determined by the inflow rate, and the gravitational and surface tension forces. Similarly, in the 
case of magnetic reconnection, the magnetic energy can be continuously supplied by inflow, build up in the vicinity 
of a flare epicentre, and, when a certain critical level of the magnetic field complexity is reached, the energy is 
released by a burst, and the buildup of the energy repeats again.  

Theoretically, a periodic regime of magnetic reconnection can be achieved in several physical situations.
One possibility is the coalescence of two magnetic flux tubes, e.g.,
a collision of two twisted coronal loops \cite{1987ApJ...321.1031T}. 
The induced oscillation is essentially anharmonic:
the magnetic field strength experiences periodic sudden increases, causing periodic spikes of the current density \cite{2009SSRv..149..119N}.
The period is estimated as a product of the plasma-$\beta$ and the transverse Alfv\'en transit time across
the region of the interaction (which is definitely smaller than the minor radii of the colliding loops). 
More generally, the period is determined by the plasma $\beta$, the magnetic twist (the ratio between the poloidal 
and toroidal components of the field in the loops), and the colliding velocity of the loops. According to numerical 
simulations, the oscillations can have double or triple sub-peaks, which is consistent with some observational 
examples of QPP \cite{1987ApJ...321.1031T}. 

During magnetic reconnection, steep gradients of the magnetic field produce high values of the electric current density, which can reach the threshold of plasma micro-instabilities and hence cause the switching from
classical to anomalous resistivity. The positive feedback between reconnection, plasma acceleration and rise of 
the resistivity can result in a periodic regime of magnetic reconnection \cite{Kliem00}. The periodicity is 
manifested through the repeated generation of magnetic islands or plasmoids, and their coalescence. 
The principle effect that leads 
to the repetitive reconnection is the inability of the plasma to carry a sufficient amount of magnetic flux into the 
diffusion region to support the Alfv\'enic outflow in the steady Petchek state. Thus the system keeps switching 
between Petchek and Sweet--Parker regimes.  The period of the repetition was found to be about thirteen Alfv\'en 
crossing times between neighbouring magnetic X-points in the generated chain of plasmoids. 
Generation of similar sequences (or chains) of plasmoids in a reconnecting current sheet  has also been 
observed in the 2D numerical resistive-MHD experiments with large Lundquist numbers  
\cite{2006PhPl...13k2105A,2009PhRvL.103j5004S}. This effect can explain 
the oscillatory reconnection in the numerical experiments on the emergence of a magnetic flux
rope into the solar atmosphere endowed with a vertical magnetic field \cite{murray09}. 
A series of reconnection reversals with the period of several minutes was observed, 
whereby reconnection occurred in distinct bursts and the inflow 
and outflow magnetic fields of one burst became the outflow and inflow fields in the following 
burst, respectively. During each burst the gas pressure in the bounded outflow regions increases 
above the level of that in the inflow regions and, eventually, gives rise to a reconnection reversal. 
 
Nonlinear coupling of the tearing mode and the Kelvin--Helmholtz instabilities was shown to have a periodic 
(or over-stable) regime \cite{2006ApJ...644L.149O} too. The most preferable conditions for this regime appear
in high $\beta$ plasmas. In 2.5D compressible visco-resistive MHD simulations, the period of the 
induced oscillations was found to be about 50 Alfv\'en transit times across the current sheet half-width.

The vicinity of a magnetic X-point, one of the preferable sites for magnetic reconnection, is a fast magnetoacoustic
resonator. Standing fast waves trapped in the resonator modify the magnetic field and plasma
density, and hence affect  periodically the reconnection rate. 
The oscillatory regime of reconnection, caused by an $m=2$ oscillation in a 2D X-type neutral magnetic
point, has been found in the numerical resistive-MHD simulations \cite{2009A&A...493..227M}. 
The nonlinear compressible fast-mode oscillation deforms the X-point, 
causing periodic creation of reconnecting current sheets. 
The plane of the sheets alternates between vertical and horizontal orientations. The principle reason for the 
oscillation is the inertial overshooting of the plasma, which carries more flux through the neutral point
than is required for static equilibrium \cite{1991ApJ...371L..41C}.

\subsection{Periodically triggered reconnection} 

In terms of the \lq\lq dripping" model discussed in Section~\ref{sec:spon}, the periodically triggered regime
of reconnection can be seen as periodic shaking of the ceiling. In this case, there is a certain region on the
parametric plane showing the period and the amplitude of the shaking force, which corresponds to the 
dripping rate coinciding with the period of the shaking. In MHD, the external shaking
can be associated with an MHD oscillation or wave generated outside the flare epicentre. 
The link of the MHD wave and the reconnection rate can be achieved by several mechanisms.

One option is if there is an oscillating plasma structure near the potential reconnection site
\cite{2005A&A...440L..59F}. Transverse oscillations
create periodic fast magnetoacoustic wave, approaching the reconnection site. The interaction of the fast wave
with, e.g., a magnetic X-point, is accompanied with a periodic creation of very localised and sharp spikes 
of the electric current density \cite{2004A&A...420.1129M}. The spike acts as a seed for the 
onset of current-driven plasma micro-instabilities, 
causing a sudden increase in the plasma resistivity in the vicinity of the X-point and hence 
periodically triggering magnetic reconnection \cite{2006A&A...452..343N}.
  
Another mechanism, proposed in \cite{2006SoPh..238..313C}, 
links the periodic triggering of reconnection by a slow magnetoacoustic wave: the
wave periodically perturb the plasma density in the reconnection site, modulating its rate.
Such a relationship between three-minute oscillations in sunspot atmospheres, interpreted as slow waves, 
and QPP in solar flares in an adjacent coronal active region was
observationally established in \cite{2009A&A...505..791S}. 

\section{Method of empirical mode decomposition}
\label{sec:emd}

\begin{figure}
  \begin{center}
  \includegraphics[scale=0.6]{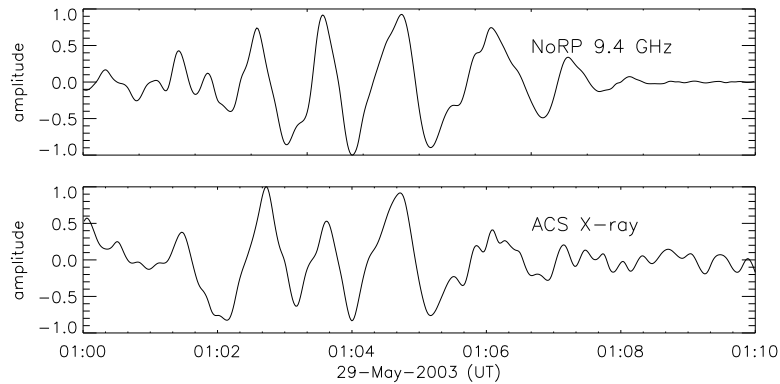}
  \end{center}
  \caption{Oscillatory components of the flare on 29 May 2003 in the microwave flux at 9.4~GHz, measured with the NoRP (upper panel) and in the hard X-ray flux measured with ACS (lower panel). The components were obtained by summing up three intrinsic modes of the signals, determined with the Empirical Mode Decomposition technique.}
\label{fig:2}
\end{figure}

The complexity of the physics associated with the generation of flaring QPP by oscillatory regimes
of magnetic reconnection, discussed above, 
makes the flaring light curves intrinsically non-stationary and nonlinear. 
Hence, proper analysis of light curves exhibiting QPP should take into account that feature. 
The traditional approach to the detection of QPP is based upon various
versions of Fourier analysis, or on wavelet analysis with the Morlet mother function (e.g. 
\cite{2009A&A...505..791S}), which both consider the signal as a sum of linear harmonic functions. 
Harmonic functions, being the eigenfunctions of linear oscillatory systems, are definitely not
the most suitable basis for the decomposition of nonlinear and non-stationary series and revealing
the physical mechanisms responsible for their time-variability.  A recently designed method of Empirical 
Mode Decomposition (EMD, \cite{1998RSPSA.454..903E}) is a promising alternative to the Fourier-based 
techniques in the study of flaring QPP. 

EMD decomposes a signal into a small number of intrinsic mode functions. 
An intrinsic mode function satisfies the following conditions: in the whole data set, the number
of extrema and the number of zero crossings must either equal or differ at most by one, and at any data point, 
the mean value of the envelope defined using the local maxima and the envelope defined using the local minima is 
zero. Thus, EMD does not require the intrinsic mode functions to be harmonic, allowing for the extraction of
signals of an anharmonic shape (e.g. such as saw-tooth, square, triangular). In application to solar coronal oscillations, the first use of EMD was in \cite{2004ApJ...614..435T}. The first application of EMD to solar flare light curves was in \cite{Inglis_thesis}.

Figure~\ref{fig:2} shows the EMD-filtering of the light curves of the flare shown in Figure~\ref{fig:l}.
EMD determined eight intrinsic modes. After removal of the apparent high-frequency noise and the long-durational 
trend, the synthesised signal keeps the anharmonicity of the oscillations, pointed out 
in Section~\ref{sec:obs}. Similarly, EMD  can be used as an adaptive 
filtering that naturally preserves the period and amplitude modulation in the signal. 

\section{Conclusions}

Observational properties of solar flare QPP, the modulation depth, the anharmonic shape and the period,
suggest their auto-oscillatory nature and the possible relationship with oscillatory regimes of magnetic
reconnection. Those regimes have been observed in several numerical MHD experiments, and can occur
spontaneously or be periodically triggered by an externally generated MHD oscillation or wave. The
study of solar and stellar flare QPP opens up very interesting perspectives for revealing the basic 
physical processes operating in flares and the mechanisms for magnetic reconnection, charged-particle acceleration
and energy deposition. An interesting avenue in the theoretical investigation would be creation of a
low-dimensional model linking observable parameters of QPP with the physical conditions in the 
reconnection site. In particular, it is necessary to understand whether the oscillatory regime is
associated with relaxation oscillations, or is better described by some kind of a limit-cycle dynamics.
Such a model would become a basis for plasma diagnostics with the use of QPP.
In data analysis, a prospective novel approach is the application of the EMD technique that allows one to
extract nonlinear and non-stationary properties of QPP. The primary questions to be answered observationally
are whether QPP is an intrinsic feature of flaring energy releases, what the typical shape of the oscillatory
pattern is, the phenomenological link between the main periods and other observables, the existence of multi-periodic
regimes and the typical period ratios. All this makes the study of solar flare QPP an interesting, 
promising and rapidly developing novel research area, which will shed light not only on the
physical processes operating in the solar and stellar atmospheres, but also of fundamental importance for
basic plasma physics.

 \ack 
The work was supported by the Royal Society UK-Russian
International Joint Project, STFC, and the grants RFBR 10-02-01285-a and 10-02-00153-a.

  \end{document}